\documentclass[conference]{IEEEtran}

\usepackage{cite}

\usepackage{url}
\usepackage{hyperref}
\usepackage[keeplastbox]{flushend}
\usepackage{xspace}

\usepackage{graphicx}
\usepackage{textcomp}

\usepackage{booktabs}
\usepackage{multirow}
\usepackage{threeparttable}
\usepackage[table]{xcolor}

\usepackage{amsmath,amssymb,amsfonts}
\usepackage{algorithm}
\usepackage{algorithmicx}
\usepackage{algpseudocode}
\makeatletter
\newcommand\fs@betterruled{%
  \def\@fs@cfont{\bfseries}\let\@fs@capt\floatc@ruled
  \def\@fs@pre{\vspace*{5pt}\hrule height.8pt depth0pt \kern2pt}%
  \def\@fs@post{\kern2pt\hrule\relax}%
  \def\@fs@mid{\kern2pt\hrule\kern2pt}%
  \let\@fs@iftopcapt\iftrue}
\floatstyle{betterruled}
\restylefloat{algorithm}
\makeatother

\usepackage{tcolorbox,tikz}
\usepackage{changepage}
\usepackage{pgfplots}
\usepackage{pgfplotstable}
\usepackage{filecontents}
\pgfplotsset{compat=1.13}
\usepgfplotslibrary{statistics}

\AtBeginDocument{%
}

\newcommand{\bears}{\textsc{Bears}\xspace}
\newcommand{\bearsCollector}{\textsc{Bears-collector}\xspace}
\newcommand{\bearsBenchmark}{\textsc{Bears-benchmark}\xspace}

\newcommand{\nbBearsBugs}{251\xspace}
\newcommand{\nbBearsProjects}{72\xspace}

\begin{document}

\title{\bears: An Extensible Java Bug Benchmark for Automatic Program Repair Studies}

\author{\IEEEauthorblockN{Fernanda~Madeiral\IEEEauthorrefmark{1},
		Simon~Urli\IEEEauthorrefmark{2},
        Marcelo~Maia\IEEEauthorrefmark{1},
        and~Martin~Monperrus\IEEEauthorrefmark{3}}
\IEEEauthorblockA{\IEEEauthorrefmark{1}Federal University of Uberl\^andia, Brazil, \{fernanda.madeiral, marcelo.maia\}@ufu.br}
\IEEEauthorblockA{\IEEEauthorrefmark{2}INRIA \& University of Lille, France, simon.urli@inria.fr}
\IEEEauthorblockA{\IEEEauthorrefmark{3}KTH Royal Institute of Technology, Sweden, martin.monperrus@csc.kth.se}
}


\maketitle

\begin{abstract}
Benchmarks of bugs are essential to empirically evaluate automatic program repair tools.
In this paper, we present \bears, a project for collecting and storing bugs into an extensible bug \underline{be}nchmark for \underline{a}utomatic \underline{r}epair \underline{s}tudies in Java.
The collection of bugs relies on commit building state from Continuous Integration (CI) to find potential pairs of buggy and patched program versions from open-source projects hosted on GitHub.
Each pair of program versions passes through a pipeline where an attempt of reproducing a bug and its patch is performed. The core step of the reproduction pipeline is the execution of the test suite of the program on both program versions. If a test failure is found in the buggy program version candidate and no test failure is found in its patched program version candidate, a bug and its patch were successfully reproduced.
The uniqueness of Bears is the usage of CI (builds) to identify buggy and patched program version candidates, which has been widely adopted in the last years in open-source projects. This approach allows us to collect bugs from a diversity of projects beyond mature projects that use bug tracking systems.
Moreover, \bears was designed to be publicly available and to be easily extensible by the research community through automatic creation of branches with bugs in a given GitHub repository, which can be used for pull requests in the \bears repository.
We present in this paper the approach employed by \bears, and we deliver the version 1.0 of \bears, which contains \nbBearsBugs reproducible bugs collected from \nbBearsProjects projects that use the Travis CI and Maven build environment.
\end{abstract}

\section{Introduction}

Automatic program repair is a recent and active research field that consists of automatically finding solutions to software bugs, without human intervention~\cite{Monperrus2018}. To measure the repairability potential of program repair tools, empirical evaluation is conducted by running them on \textit{known bugs}. Each known bug consists of a buggy program version and a mechanism for exposing the bug, such as a failing test case. Some evaluations are ad-hoc, but the most valuable ones from a scientific viewpoint are based on \textit{benchmarks of bugs}  constructed using systematic approaches.

In 2015, Le Goues et al.~\cite{LeGoues2015} claimed that ``Since 2009, research in automated program repair [...] has grown to the point that it would benefit from carefully constructed benchmarks'', which motivates research towards the construction of well-designed benchmarks of bugs. A well-designed benchmark  1) simplifies experimental reproduction; 2) helps to ensure generality of results; 3) allows direct comparisons between competing methods; and 4) enables measurement of a research field progress over time~\cite{LeGoues2015}.

Collecting bugs, however, is a challenging task. Le Goues et al.~\cite{LeGoues2013} highlighted that a key challenge they faced was  finding a good benchmark of bugs to evaluate their repair tool. They also claimed that a good benchmark should include \textit{real} and easy to \textit{reproduce} bugs from a variety of real-world systems. Durieux et al.~\cite{Durieux2017-NPEFix} also pointed out that creating a benchmark of bugs is challenging. They reported that it is difficult to \textit{reproduce failures}, and it can take up to one day to find and reproduce a single null pointer exception bug.

The seminal benchmarks of bugs dedicated for automatic program repair studies were for C programs, namely ManyBugs and IntroClass~\cite{LeGoues2015}. For the Java language, Defects4J~\cite{Just2014} (395 bugs from six projects) has been intensively used to evaluate state-of-the-art repair tools (e.g. jGenProg~\cite{Martinez2016-Astor} and Nopol~\cite{Xuan2016-Nopol}). More recently, Bugs.jar \cite{Saha2018} (1,158 bugs from eight projects) was also proposed, increasing the number of Java bugs available for the automatic program repair community.

Both Defects4J and Bugs.jar are based on the same construction strategy, consisting of going through past commits in version control systems, with the support of bug trackers to find bug fixing commits.
However, they include bugs from only 13 projects in total (one project is in common), which are all \textit{mature} ones.
This is a major threat, because bugs in benchmarks should come from a representative sample of real-world projects, considering diversity in several aspects.

Moreover, the current benchmarks are barely updated, if so. For instance, since its creation in 2014, Defects4J has evolved only once, where 38 bugs collected from one single project were included in the dataset. Since new projects, new developing practices, and new language features are proposed over time, new releases of a benchmark are desirable to keep it relevant. To do so, the benchmark must be extensible upfront, with a modular and reusable design.

To address those problems, we present \bears, a project for collecting and storing bugs into an extensible bug \underline{be}nchmark for \underline{a}utomatic \underline{r}epair \underline{s}tudies in Java. \bears is divided in two components: the \bearsCollector and the \bearsBenchmark. The former is the tool for collecting and storing bugs, and the latter is the actual benchmark of bugs.

Differently from Defects4J and Bugs.jar, the approach employed by the \bearsCollector to automatically identify bugs and their patches is based on Continuous Integration (CI) builds. One of the main phases of a CI build is the testing phase~\cite{Beller2017}. Our idea is that the statuses of CI builds indicate potential program versions containing bugs and their patches that are compilable and testable, which is a requirement towards a reproducible bug.

Moreover, \bears uses GitHub in an original way in order to support the benchmark evolution. The \bearsCollector automatically creates a publicly available branch in a given GitHub repository for each successfully reproduced bug. One can extend \bears by opening pull requests on the main repository of the \bearsBenchmark.

To sum up, our contributions are:

\begin{itemize}
\item A novel concept: we present an original approach to collect bugs and their patches, based on commit building state from Continuous Integration;
\item A tool, called \bearsCollector: it implements our approach, focusing on Java, Travis CI and Maven projects, and it is publicly available for future research;
\item A benchmark of bugs, called \bearsBenchmark: our benchmark contains \nbBearsBugs bugs from \nbBearsProjects projects; to our knowledge, it is the largest benchmark of reproducible bugs with respect to project diversity (the closest benchmark is Bugs.jar, which covers only eight projects).
\end{itemize}

The remainder of this paper is organized as follows.
\autoref{sec:bears-design-decisions} presents the design decisions at conceiving \bears.
\autoref{sec:bears-bug-collection-process} presents the process of the \bearsCollector, which is used in three execution rounds to collect bugs in \autoref{sec:collecting-bugs}.
The collected bugs compose the first version of the \bearsBenchmark, which is presented in \autoref{sec:content-of-bears}.
\autoref{sec:discussion} presents challenges, limitations, and threats to validity.
Finally, \autoref{sec:related-work} presents the related works, and
\autoref{sec:conclusion} presents the final remarks and the open-science artifacts produced in this work.

\section{\bears Design Decisions}\label{sec:bears-design-decisions}

\begin{figure*}[ht]
  \centering
  \includegraphics[width=\textwidth]{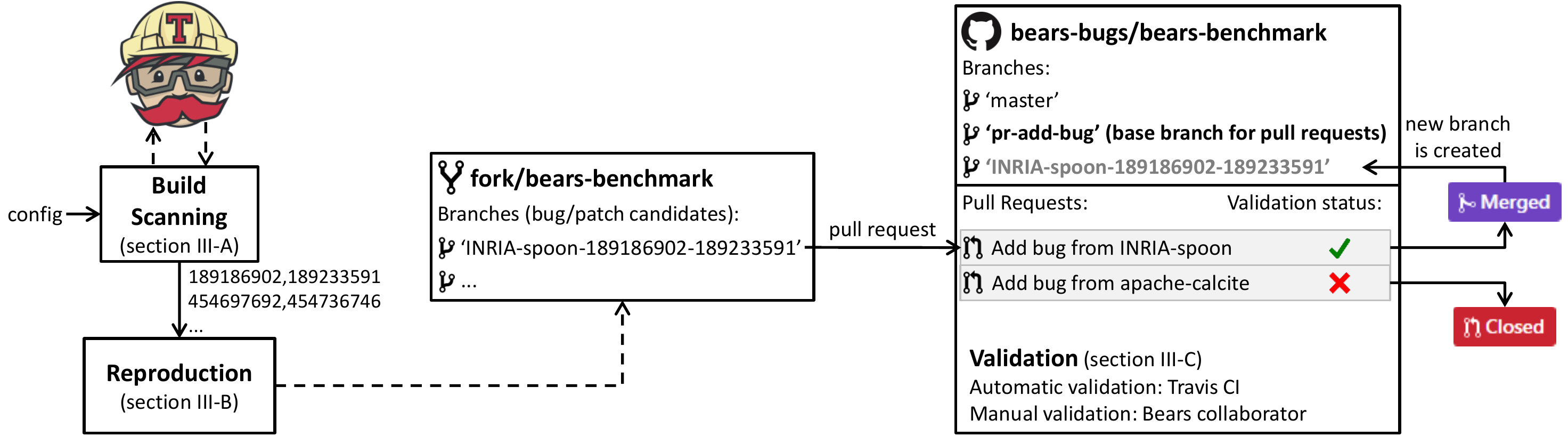}
  \caption{Overview of the \bears process: its uniqueness is to be based on commit building state from Travis CI.}
  \label{fig:bears}
\end{figure*}

In this section, we present the main design decisions for \bears: the intuition, the bug identification strategy based on CI build statuses, the criteria that projects must meet to be used as source for collecting bugs with the \bearsCollector, the criteria that bugs must meet to be included in the \bearsBenchmark, and the public bug repository design.

\subsection{Bug Collection based on Continuous Integration}

The closest related works to \bears (Defects4J~\cite{Just2014} and Bugs.jar \cite{Saha2018}) are based on mining past commits.
In \bears, we radically depart from this technique and explore an original path: we use continuous integration to create our benchmark.
Our idea is that the statuses of CI builds (failed, errored and passed) can guide us at finding 1) program versions that contain test failures and 2) their ground-truth patches written by developers.

Travis CI is a CI platform that tightly integrates with GitHub. It has emerged as the most popular and used CI platform for open-source software development, and it has recently started to get attention of researchers as a data source to conduct research: the mining challenge at MSR'17 (The 14th International Conference on Mining Software Repositories) was on a dataset synthesized from Travis CI and GitHub~\cite{Beller2017-challenge}.

In Travis CI, a build marked as \textit{errored} means that some failure has happened in a phase of the build life-cycle that is before the execution of tests, thus the version of the program when such build was triggered is not interesting for us (we focus on test failures). On the other hand, builds marked as \textit{passed} suggest compilable and testable program versions.

\subsection{Buggy and Patched Program Versions from CI Builds}\label{sec:build-pair-cases}

In CI, each build ($b$) is associated with a program version (by its commit)\footnote{To refer to specific CI builds in this paper, we use their IDs with hyperlink to the them (in Travis). Last accessed to these links: December 13, 2018.}.
Our build-based approach identifies pairs of buggy and patched program versions from pairs of immediately subsequent CI builds, $b_{n}$ and $b_{n+1}$. There are two cases where a given pair ($b_{n}$, $b_{n+1}$) may contain a buggy and a patched program version, ($b_{buggy}$, $b_{patched}$), based on the CI build statuses and the files changed between the pair of builds:

\vspace{5pt}
\noindent\textit{Case \#1: failing-passing builds with no test changes}

\noindent\textit{Definition:}
$b_{n}$ is a failing build and $b_{n+1}$ is a passing build that does not contain changes in test files, and $b_{n}$ fails because at least one test case fails.

\noindent\textit{Example:}
Consider the pair of builds ($b_{n}$=\href{https://travis-ci.org/CorfuDB/CorfuDB/builds/330246430}{330246430}, $b_{n+1}$=\href{https://travis-ci.org/CorfuDB/CorfuDB/builds/330267605}{330267605}) from the CorfuDB/CorfuDB project.
$b_{n}$ failed in Travis CI because one test case failed due to \texttt{ComparisonFailure}, and $b_{n+1}$ passed, where the message of the commit associated with it is ``fix test''.

\vspace{5pt}
\noindent\textit{Case \#2: passing-passing builds with test changes}

\noindent\textit{Definition:}
$b_{n}$ is a passing build and $b_{n+1}$ is also a passing build, but $b_{n+1}$ contains changes in test files, and at least one test case fails when the source code from $b_{n}$ is tested with the test cases from $b_{n+1}$.

\noindent\textit{Example:}
Consider the pair of builds ($b_{n}$=\href{https://travis-ci.org/spring-cloud/spring-cloud-gcp/builds/330973656}{330973656}, $b_{n+1}$=\href{https://travis-ci.org/spring-cloud/spring-cloud-gcp/builds/330980388}{330980388}) from the spring-cloud/spring-cloud-gcp project. Both builds passed in Travis CI. However, test files were changed in the commit that triggered the second passing build. The source code from $b_{n}$, when tested with the tests from $b_{n+1}$, fails due to \texttt{NullPointerException}. Indeed, the commit message from $b_{n+1}$, ``Fix NPE for GCS buckets that have underscores'', confirms that a null-pointer exception had been fixed.

\vspace{5pt}
The difference between both cases is about the development practice.
In case \#1, the bug-triggering test is contained in $b_{buggy}$, which makes $b_{n}$ fail, and in case \#2, the bug-triggering test is contained in $b_{patched}$ together with the patch, which makes $b_{n}$ pass since there is no bug-triggering test on it.

\subsection{Inclusion Criteria for Projects}\label{sec:inclusion-criteria-for-projects}

The criteria that a project must meet to be considered by the \bearsCollector are the following: 
1) the project must be publicly available on GitHub, 
2) it must use Travis continuous integration service, and 
3) it must be a Maven project. 
These three conditions are required to keep engineering effort reasonable (supporting different code hosting, CI and build technologies are out of the scope of a laboratory project).

\subsection{Inclusion Criteria for Bugs}\label{sec:inclusion-criteria-for-bugs}

The criteria that a bug must meet to be included in the \bearsBenchmark are the following:

\vspace{5pt}
\noindent\textit{Criterion \#1--The bug must be reproducible.} To repair a given bug, test-suite based repair tools create patches  that should make the whole test suite of the program pass, thus they need at least one failing test case that triggers the bug. Therefore, each bug contained in the \bearsBenchmark must be accompanied with at least one bug-triggering test case.

\vspace{5pt}
\noindent\textit{Criterion \#2--The bug must have been fixed by a human.} A patch generated by a test-suite based repair tool, even when it makes the whole test suite pass, may be incorrect. When manually evaluating the correctness of a patch generated by a repair tool for a given bug, one of the most valuable resources that researchers use is the human-written patch. To allow this type of manual study, which is essential for making sound progress, we ensure that each bug contained in the \bearsBenchmark is accompanied with its human-written patch.

\subsection{Bug Repository Design}\label{sec:bug-storing-design}

A well-designed bug collection process involves the important feature of storing the bugs and their patches.
We designed such feature aiming
1) to keep the bugs organized,
2) to make the bugs publicly available for the research community, and
3) to make easier for other researchers to collect more bugs using the \bearsCollector and to include them in the \bearsBenchmark.

To achieve goal \#2, we decided to automatically store bugs in a public GitHub repository, which is given as input to the \bearsCollector. To achieve goal \#1, we defined that the internal organization of such repository is based on branches, so that when a pair of builds is successfully reproduced by the \bearsCollector, a new branch is created in the given GitHub repository. Since the repository is settable on the \bearsCollector, and the result produced by the \bearsCollector is branch-based, we achieve goal \#3 by allowing pull requests in the \bearsBenchmark repository from the given GitHub repository (when it is a fork from the \bearsBenchmark repository).

The branches generated by the \bearsCollector are standardized. The name of each branch follows the pattern \texttt{\fontsize{8.5}{12}\selectfont <project slug>-<buggy build id>-<patched build id>}, and each branch contains the sequence of commits presented in~\autoref{tab:bears-commits}.
Moreover, each branch contains a special file named \texttt{bears.json}, which is a gathering of information collected during the bug reproduction process. This file is based on a json schema that we created to store key properties. It contains information about the bug (e.g. test failure names), the patch (e.g. patch size), and the bug reproduction process (e.g. duration).

\begin{table}
    \caption{The canonical commits in a \bears git branch.}
    \label{tab:bears-commits}
    \centering
    \begin{tabular}{l l l}
        \toprule
         \# & Cases & Commit content \\
        \midrule
        1 & both & the version of the program with the bug \\
        2 & case \#2 & the changes in the tests \\
        3 & both & the version of the program with the human-written patch \\
        4 & both & the metadata file \texttt{bears.json} \\
        \bottomrule
    \end{tabular}
\end{table}

\section{The \bearsCollector Process}\label{sec:bears-bug-collection-process}

The overview of the \bears process for collecting and storing bugs is illustrated in \autoref{fig:bears}. In a nutshell, given an input configuration, pairs of builds are scanned from Travis CI (\textit{Build Scanning}). Each pair of builds passes through a reproduction pipeline towards finding a test failure followed by a passing test suite, i.e. a reproducible bug and its patch (\textit{Reproduction}). Each successfully reproduced pair is stored in a dedicated branch in a GitHub repository. If such repository is a fork from the \bearsBenchmark repository, the branches can be used for opening pull requests for the addition of bugs into the \bearsBenchmark: the special branch named ``pr-add-bug'' is used as base branch. An open pull request is automatically validated by a build triggered in Travis CI, and manually validated by a collaborator of \bears (\textit{Validation}). If the pull request passes in both validations, it is merged, and a new branch for the bug is created in the \bearsBenchmark; otherwise it is closed. The three main phases of this process are presented in dedicated sections as follows.

\subsection{Phase I--Build Scanning}\label{sec:phaseI-Build-Scanning}

The first phase of the \bearsCollector process is the build scanning. The goal of this phase is to automatically scan build ids from Travis CI and select pairs of them to be reproduced (Phase II--Reproduction), such that the selected pairs correspond to one of the two cases presented in \autoref{sec:build-pair-cases}.

There are two ways that the scanning can be performed: 1) for a fixed period, or 2) in real time.
For the former, the scanner takes as input a list of projects (that meet the criteria presented in \autoref{sec:inclusion-criteria-for-projects}) and a time window (e.g. 01/01/2018--31/01/2018), and brings all builds from these projects that have finished to run in Travis during such time window.
In the latter, the scanner repeatedly checks on Travis (e.g. in each minute) which builds have finished to run, independently of a given list of projects.

In both ways, after scanning build ids, pairs of build ids are created, and then they pass through a selection process. To do so, we first gather the builds that passed in Travis out of all scanned builds, which are the $b_{patched}$ candidates. Then, each $b_{patched}$ candidate is given as input to \autoref{alg:build-scanning}, where its previous build is retrieved (line 1), the $b_{buggy}$ candidate.

The algorithm checks if the $b_{buggy}$ candidate failed in Travis (line 4). If it did, and if the $diff$ (changes) between the commits that triggered the two builds contains Java source code file, but does not contain test file (line 5), the pair ($b_{buggy}$, $b_{patched}$) is successfully returned (line 6), because such pair of builds can fit in case \#1 (failing-passing builds with no test changes). If the $b_{buggy}$ candidate passed in Travis (line 8), and if the $diff$ between the commits contains Java source code file and test file (line 9), the pair ($b_{buggy}$, $b_{patched}$) is successfully returned (line 10), because such pair of builds can fit in case \#2 (passing-passing builds with test changes). The pairs of builds collected with this algorithm compose the input for Phase II--Reproduction.

\begin{algorithm}[t]
	\caption{Assessing suitability of a scanned $b_{patched}$ candidate for reproduction.}
    \label{alg:build-scanning}
    \small
    \begin{algorithmic}[1]
      \Require $b_{patched}$ candidate
      \Ensure pair ($b_{buggy}$, $b_{patched}$) or $null$
      \State $b_{buggy}$ candidate $\leftarrow$ previous build from $b_{patched}$ candidate
      \State $c_{buggy}$ $\leftarrow$ commit that triggered $b_{buggy}$
      \State $c_{patched}$ $\leftarrow$ commit that triggered $b_{patched}$
      \If{$b_{buggy}$ is failing in Travis CI} \Comment{case \#1}
      	\If{failure in Travis is due to test failure \textbf{and} \newline
      	  \hspace*{19pt} $diff(c_{buggy}$, $c_{patched})$ contains Java source code file  \textbf{and} \newline
      	  \hspace*{19pt} $diff(c_{buggy}$, $c_{patched})$ does not contain Java test file}
      	    \State\hspace*{10pt}\Return($b_{buggy}$, $b_{patched}$)
      	\EndIf
      \Else \Comment{case \#2}
        \If{$diff(c_{buggy}$, $c_{patched})$ contains Java source code file \textbf{and} \newline
          \hspace*{22pt}$diff(c_{buggy}$, $c_{patched})$ contains Java test file}
      	    \State\hspace*{10pt}\Return($b_{buggy}$, $b_{patched}$)
      	\EndIf
      \EndIf
      \State\Return $null$
    \end{algorithmic}
\end{algorithm}

\begin{algorithm}
	\caption{Assessing reproducibility for a build pair.}
    \label{alg:reproduction}
    \small
    \begin{algorithmic}[1]
      \Require pair ($b_{buggy}$, $b_{patched}$)
      \Require $repo$: a GitHub repository where a new branch is created if the reproduction attempt succeeds
      \Ensure reproduction attempt status (it indicates success or failure of the reproduction attempt, and in case of success, a branch has been created in $repo$ by the algorithm)
      \State \textbf{var} $pipeline$: a queue with FIFO steps
      \State $pipeline$ $\leftarrow$ clone repository \Comment{`$\leftarrow$' means `append'}
      \State \textbf{$\triangleright$ Beginning of the reproduction towards a test failure}
      \If{$b_{buggy}$ is failing in Travis CI} \Comment{case \#1}
      	\State $pipeline$ $\leftarrow$ check out the commit from $b_{buggy}$
      \Else \Comment{case \#2}
        \State $pipeline$ $\leftarrow$ check out the commit from $b_{patched}$
        \State $pipeline$ $\leftarrow$ compute source code and test code directories
        \State $pipeline$ $\leftarrow$ check out only source code files from $b_{buggy}$
      \EndIf
      \State $pipeline$ $\leftarrow$ build project
      \State $pipeline$ $\leftarrow$ run tests
      \State $pipeline$ $\leftarrow$ analyze test results \Comment{\textbf{\footnotesize test failure must be found}}
      \State \textproc{InitRepository($pipeline$, $b_{buggy}$, $b_{patched}$)}
      \State \textbf{$\triangleright$ Beginning of the reproduction towards a passing test suite}
      \State $pipeline$ $\leftarrow$ check out the commit from $b_{patched}$
      \State $pipeline$ $\leftarrow$ build project
      \State $pipeline$ $\leftarrow$ run tests
      \State $pipeline$ $\leftarrow$ analyze test results \Comment{\textbf{\footnotesize no test failure must be found}}
      \State \textproc{CreateBranchInRepository($pipeline$, $repo$)}
      \For{$i\leftarrow 1$ \textbf{to} $pipeline.size$}
        \State $step$ $\leftarrow$ $pipeline[i]$
        \State $status$ $\leftarrow$ run $step$
        \If{$status$ is failed}
        	\State\Return failure status \Comment{abort pipeline execution}
        \EndIf
      \EndFor
      \State\Return success status
    \end{algorithmic}
\end{algorithm}

\begin{algorithm}
	\caption{\small\textproc{InitRepository($pipeline$, $b_{buggy}$, $b_{patched}$)}}
	\label{alg:InitRepository}
	\small
    \begin{algorithmic}[1]
        \If{$b_{buggy}$ is failing in Travis CI} \Comment{case \#1}
          \State $pipeline$ $\leftarrow$ commit files (commit \#1 in \autoref{tab:bears-commits})
        \Else \Comment{case \#2}
            \State $pipeline$ $\leftarrow$ check out only test code files from $b_{buggy}$
            \State $pipeline$ $\leftarrow$ commit files (commit \#1 in \autoref{tab:bears-commits})
            \State $pipeline$ $\leftarrow$ check out only test code files from $b_{patched}$
            \State $pipeline$ $\leftarrow$ commit files (commit \#2 in \autoref{tab:bears-commits})
        \EndIf
    \end{algorithmic}
\end{algorithm}

\begin{algorithm}[!ht]
	\caption{\small\textproc{CreateBranchInRepository($pipeline$, $repo$)}}
	\label{alg:CreateBranchInRepository}
	\small
    \begin{algorithmic}[1]
        \State $pipeline$ $\leftarrow$ commit files (commit \#3 in \autoref{tab:bears-commits})
        \State $pipeline$ $\leftarrow$ commit files (commit \#4 in \autoref{tab:bears-commits})
        \State $pipeline$ $\leftarrow$ create a new branch and push it to $repo$
    \end{algorithmic}
\end{algorithm}

\subsection{Phase II--Reproduction}\label{sec:phaseII-Reproduction}

The goal of this phase is to test if the pairs of builds obtained in Phase I--Build Scanning fit in one of the two cases presented in \autoref{sec:build-pair-cases}. To do so, we submit each pair of builds to the \textit{reproduction} process, which consists of building both program versions from the build pair, running tests, and observing the test results. If a test failure is found only in the program version from the $b_{buggy}$ candidate, but not in the one from the $b_{patched}$ candidate, a bug and its patch were reproduced.

To perform the reproduction, we designed and implemented a \textit{pipeline of steps}. \autoref{alg:reproduction} presents the core steps for a given build pair candidate ($b_{buggy}$, $b_{patched}$). This algorithm has two main phases: the pipeline construction (line 1 to 20) and the pipeline execution (line 21 to 28). The pipeline construction consists of creating a sequence of steps that should be further executed in order, which are stored in the variable $pipeline$ (line 1). The pipeline execution consists of running the $pipeline$ step by step. If any step fails, the execution of the pipeline is aborted (line 25). Otherwise, the execution of the pipeline ends with success (line 28), meaning that a test failure was found in the program version from the $b_{buggy}$ candidate, and no test failure was found in the program version from the $b_{patched}$ candidate. Note that some pipeline steps are responsible for storing data by committing and pushing to a GitHub repository, according to the design for storing bugs described in \autoref{sec:bug-storing-design}. These steps are in \autoref{alg:InitRepository} and \autoref{alg:CreateBranchInRepository}, which are called from the main reproduction algorithm (\autoref{alg:reproduction}) in lines 14 and 20. We separated these steps from the main algorithm to make easier to understand the core steps of the reproduction process presented in \autoref{alg:reproduction}, but yet providing the completeness of the process.

The general idea of \autoref{alg:reproduction} is to try to reproduce a test failure from the source code of the $b_{buggy}$ candidate first, and then a passing test suite from the source code of the $b_{patched}$ candidate. To do so, the algorithm first makes a local copy of the remote GitHub repository where the pair of builds were triggered from (line 2). Then, the algorithm checks out the repository to the point of interest in the project history. If the $b_{buggy}$ candidate failed in Travis (i.e. the pair of builds is in case \#1), the algorithm checks out the commit that triggered $b_{buggy}$ (line 5), in order to test the source code from $b_{buggy}$ with the tests from $b_{buggy}$. If the $b_{buggy}$ candidate passed in Travis (i.e. the pair of builds is in case \#2), the algorithm first checks out the commit that triggered $b_{patched}$ (line 7), which contains the tests to be executed in the source code from $b_{buggy}$. Then, the computation of source code and test code directories is performed (line 8), so that only the source code files from the commit that triggered $b_{buggy}$ are checked out (line 9), in order to test the source code from $b_{buggy}$ with the tests from $b_{patched}$.

After the checking out steps, the algorithm builds the project (line 11), runs tests (line 12), and analyzes the test results (line 13). Since the $b_{buggy}$ candidate is being tested, at least one test case must fail. If it does happen, it means that the algorithm found a reproducible bug, and its execution continues in order to try the reproduction of a passing test suite from the $b_{patched}$ candidate. Then, it checks out the commit that triggered $b_{patched}$ (line 16), builds the project (line 17), runs tests (line 18), and analyzes the test results (line 19). Since the $b_{patched}$ candidate is being tested, all tests must pass. If it does happen, it means that the algorithm reproduced the patch for the bug. At the end of a successful reproduction attempt, a branch is created in a GitHub repository given as input to the algorithm, according to the design presented in \autoref{sec:bug-storing-design}.

\subsection{Phase III--Validation}\label{sec:phaseIII-Validation}

To include a branch generated in Phase II--Reproduction in the \bearsBenchmark repository, a pull request should be created with such branch (see \autoref{fig:bears}), using the special branch ``pr-add-bug'' as base branch. Once a pull request is open, it is validated in two steps:

\vspace{5pt}
\noindent\textit{Automated validation.} The creation of the pull request triggers a build in Travis, which runs a set of scripts to check if the content in the pull request (from the branch) is valid. One checking script is, for instance, that the commit with the buggy program version (commit \#1 in \autoref{tab:bears-commits}) has test failures when the tests are executed on it.

\vspace{5pt}
\noindent\textit{Manual validation.} A collaborator of \bears performs a manual analysis of the pull request to check if the proposed branch contains a genuine bug. It might include, for instance, the analysis of the buggy and patched source code diff and also the test failures (contained in the \texttt{bears.json} file).

\vspace{5pt}
A pull request that passed in both validations is merged, and a new branch is automatically created in the \bearsBenchmark repository with the content of the pull request; otherwise, it is closed.

\subsection{Implementation}

The main tools used in the implementation of the \bearsCollector are the following.
In Phase I--Scanning, we rely on jtravis \cite{jtravis}, a Java API to use the Travis CI API.
In Phase II--Reproduction, we use different tools depending on the pipeline step. The step \textit{clone repository} and the \textit{check out} ones are performed using JGit \cite{JGit} 
and also by directly invoking \texttt{git} commands. Since we are working with Maven projects, we use Apache Maven Invoker \cite{ApacheMavenInvoker} to invoke Maven in the steps \textit{build project} and \textit{run tests}, and we use Maven Surefire Report Plugin \cite{MavenSurefireReportPlugin} to gather test information in the step \textit{analyze test results}. To execute the pipeline, we use Docker containers, which are based in a Docker image configured to use JDK-8 and Maven 3.3.9.
Finally, in Phase III--Validation, we rely on Travis to run the automatic validation.

\section{Report on Bug Collection with the \bearsCollector}\label{sec:collecting-bugs}

To create the first version of the \bearsBenchmark, we used the \bearsCollector process described in the previous section to collect bugs.
We performed three execution rounds: two rounds are from a given time window and a list of projects, and the third one is from real time scanning of builds. In this section, we present and discuss our results organized along the three phases of the bug collection process.

\subsection{Scanning 168,772 Travis Builds}

In the scanning phase, the \bearsCollector scans and selects pairs of builds from Travis CI to be reproduced (see \autoref{sec:phaseI-Build-Scanning}). We used the following input configuration for each execution round:

\pagebreak
\noindent\textit{Execution round \#1 (time window Jan.--Dec. 2017):} we queried the most popular Java projects based on the number of watchers on GHTorrent dataset \cite{GHTorrent}. Then, we filtered out the projects that do not meet the criteria presented in~\autoref{sec:inclusion-criteria-for-projects}. It reduced the list of projects to 1,611 projects. This list was used in early experimentation during the development of the \bearsCollector. We then selected four projects that resulted in a high number of successfully reproduced build pairs in the early experimentation: INRIA/spoon, traccar/traccar, FasterXML/jackson-databind, and spring-projects/spring-data-commons. We set up the scanning of the execution round \#1 with these four projects, to comprehensively scan all builds from January to December 2017.

\vspace{5pt}
\noindent\textit{Execution round \#2 (time window Jan.--Apr. 2018):} we queried the top-100 projects that could be reproduced in the Repairnator project~\cite{Urli2018}, which were used for the scanning of the execution round \#2 from January to April 2018.

\vspace{5pt}
\noindent\textit{Execution round \#3 (real time):} we also scanned pairs of builds in real time. By doing so, there is no need to give a list of projects as input to the \bearsCollector, since the scanner directly fetches builds from Travis in real time. We collected data with the execution round \#3 within around 2 months in 2018 (Sep/19--Sep/24 and Sep/30--Nov/30).

\vspace{5pt}
\autoref{tab:scanning} summarizes the input and the scanning results for each execution round. In total, we scanned 168,772 builds over one year and a half. From these scanned builds, we obtained 12,355 pairs of builds from \autoref{alg:build-scanning}, where 741 pairs are from case \#1 (failing-passing builds), and 11,614 pairs are from case \#2 (passing-passing builds).

\begin{table}[h]
    \caption{Scanning results over three rounds.}
    \label{tab:scanning}
    \centering
    \begin{tabular}{p{.15\textwidth} r r r r}
        \toprule
         & \multicolumn{3}{c}{Execution Round} & \multirow{2}{*}{Total} \\
         & \multicolumn{1}{c}{\#1} & \multicolumn{1}{c}{\#2} & \multicolumn{1}{c}{\#3} & \\
        \midrule
        Period & 1 year & 4 months & $\sim$2 months &  \\
        \# Input projects & 4 & 100 & -- &  \\
        \# Total scanned builds & 4,987 & 66,621 & 97,164 & 168,772 \\
        \# Build pairs in case \#1 & 17 & 590 & 134 & 741 \\
        \# Build pairs in case \#2 & 1,027 & 7,755 & 2,832 & 11,614 \\
        \# Total build pairs & 1,044 & 8,345 & 2,966 & 12,355 \\
        \bottomrule
    \end{tabular}
\end{table}

There is a large difference between the number of pairs in case \#1  (failing-passing builds) and the number of pairs in case \#2  (passing-passing builds): only 6\% of the total build pairs are from case \#1.
This suggests that developers do not usually break builds by test failure, or if they do, they fix the test failure in the next build by also changing the test source code. Note that, in \autoref{alg:build-scanning}, we only accept pairs in case \#1 for reproduction that there is no test change between the builds. Our idea is that the failure that happened in Travis is fixed in the next build by only changing the source code.
Moreover, since both builds passed in Travis in case \#2, but there is test change between them, such case potentially includes pairs where feature addition and refactoring were performed, which explains the higher number of pairs in case \#2 over case \#1.


\subsection{Reproducing 12,355 Build Pairs}

The 12,355 pairs of builds obtained in the scanning phase were all analyzed through the reproduction pipeline presented in~\autoref{alg:reproduction} (see \autoref{sec:phaseII-Reproduction}). Each reproduction attempt ends with a status, which either indicates success or failure of the reproduction itself. \autoref{fig:statuses-reproduction-attempts} shows the distribution of the 12,355 reproductions by status. We had 856 successful reproductions that resulted in branches, which is 7\% of the reproduction attempts. We report on the failure statuses of the 93\% failed reproduction attempts as following.

\pgfplotstableread[col sep=comma, header=true]{status.csv}{\datatable}
\pgfplotstablegetrowsof{\datatable}
\edef\numberofrows{\pgfplotsretval}

\pgfplotsset{%
    discard if not/.style 2 args={
        x filter/.code={
            \edef\tempa{\thisrow{#1}}
            \edef\tempb{#2}
            \ifx\tempa\tempb
            \else
                \def\pgfmathresult{inf}
            \fi
        }
    }
}

\begin{figure}[h]
  \centering
  \scriptsize
  \begin{tikzpicture}
    \begin{axis}
    [xbar, 
    width=0.96\linewidth,
    height=6.8cm,
    bar width=8pt,
    visualization depends on={x \as \originalvalue},
    point meta={x/12355*100},
    nodes near coords={\pgfmathprintnumber{\originalvalue}~~\pgfmathprintnumber[fixed,precision=1]{\pgfplotspointmeta}\%},
    nodes near coords align={horizontal},
    xlabel=\# Pairs of builds,
    xlabel near ticks,
    xmin=0,
    xmax=6700,
    ytick={0,1,...,\numberofrows},
    yticklabels from table={\datatable}{outcome},
    yticklabel style={align=right, text width=1.6cm},
    y dir=reverse,
    enlarge y limits=0.08,
    xtick pos=left,
    ytick pos=left,
    legend pos=south east,
    legend cell align={left},
    legend image code/.code={
        \draw [#1] (0cm,-0.1cm) rectangle (0.3cm,0.07cm);
    },
    ]
    \addplot[bar shift=0pt, draw=black, fill=black!40!green] table[discard if not={keyword}{S}, x=sum, y expr=\coordindex, col sep=comma]{status.csv};
    \addplot[bar shift=0pt, draw=black, fill=black!30!red] table[discard if not={keyword}{F}, x=sum, y expr=\coordindex, col sep=comma]{status.csv};
    \legend{Successful reproduction, Failed reproduction}
    \end{axis}
  \end{tikzpicture}
  \vspace{-12pt}
  \caption{Statuses of the reproduction attempts.}
  \label{fig:statuses-reproduction-attempts}
\end{figure}
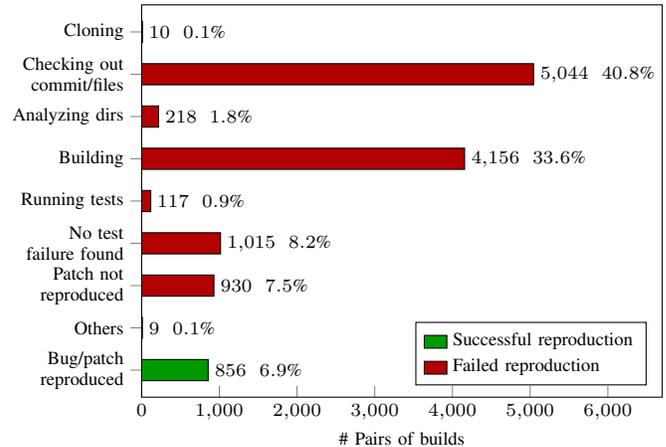

\vspace{5pt}
\noindent\textit{Failure when cloning.} In the execution round \#3 (real time), we had ten reproduction attempts that failed in the first reproduction pipeline step, i.e. when cloning the remote GitHub repository for a build pair. All of these occurrences happened with build pairs from the same project. 

\vspace{5pt}
\noindent\textit{Failure when checking out.} The most frequent failure in the reproduction attempts, occurring in 40.8\% of all attempts, is due to the checkout of commits/files. This failure happens when the commit is missing from the history of the remote repository. A missing commit happens when it was directly deleted or when the corresponding branch was deleted. The former might have happened in several ways with advanced usage of Git (e.g. with `git commit --amend' and `git reset' commands). The latter usually happens when developers delete branches used in merged pull requests.

\vspace{5pt}
\noindent\textit{Failure when analyzing directories.} For pairs of builds from case \#2, there is a pipeline step that searches for the directories containing source code and test code. In a few cases (1.8\%), this step did not succeed to find the two types of directories.

\vspace{5pt}
\noindent\textit{Failure when building.} 33.6\% of the reproduction attempts failed when building (including compiling) the project. This is due to three main reasons. First, some dependencies were missing. This happens, for instance, when dependencies are not declared in the \texttt{pom.xml} file of the project. Second, due to case \#2, when the source code and test code files from two different commits are mixed, compilation errors naturally happen. Third, we set up a timeout for each Maven goal used in the pipeline: if no output is received from Maven in 10 minutes (same value as Travis CI), the pipeline is aborted.

\vspace{5pt}
\noindent\textit{Failure when running tests.} In a few cases (0.9\%), reproduction attempts failed during the testing step. We observed that this also happens due to the timeout on Maven goal.

\vspace{5pt}
\noindent\textit{No test failure found.} The first program version (buggy version candidate) being reproduced is supposed to result in test failure(s) when the test suite is executed. In 8.2\% of the reproduction attempts, the entire test suite passed, so the reproduction was aborted. This naturally happens in case \#2, since both builds passed in Travis, which means that the changes in the tests between the two program versions do not trigger any bug. In case \#1, this can be explained by flaky tests \cite{Luo2014}, with test failures happening in Travis but not locally.

\vspace{5pt}
\noindent\textit{Patch not reproduced.} The second program version (patched version candidate) being reproduced is supposed to result in a complete passing test suite. However, this does not happen in 7.5\% of the reproduction attempts. As in \textit{no test failure found} status, flaky tests might also be one of the reasons why test failures happened locally but not in Travis. Moreover, the environment used locally might be different from the one used in Travis (e.g. Java version), leading to different test results.

\vspace{5pt}
\noindent\textit{Other issues.} Nine reproduction attempts failed due to other five different reasons. First, we had three reproduction attempts that stayed running up to three hours, constantly generating output, so the timeout with no output was not reached. We forced these reproductions to stop. Second, two reproduction attempts crashed in committing data steps (see \autoref{alg:InitRepository} and \autoref{alg:CreateBranchInRepository}) due to the lack of disk space. Third, two reproduction attempts did not succeed to push a branch (last step of the pipeline). Forth, one reproduction attempt was aborted in the very beginning (before cloning the repository) due to network issue. Finally, one reproduction attempt crashed due to an uncaught exception when gathering test information with Maven Surefire Report Plugin.

\subsection{Validating 856 Branches}\label{sec:validating-branches}

The 856 branches generated in the reproduction phase passed through the two-step validation phase (see \autoref{sec:phaseIII-Validation}).
The sequence of conducting the automatic and manual validation does not change the result: only branches that are considered valid in both validations are included in the \bearsBenchmark.
For the first version of the \bearsBenchmark, the first author of this paper started by the manual validation to have a complete picture on the obtained branches, and then the valid ones were submitted to the automatic validation\footnote{For future studies, we suggest the conduction of the automatic validation before the manual validation in order to minimize manual effort.}.

Out of 856, 295 branches were considered valid in the manual validation, and \nbBearsBugs of them were successfully validated by the automatic validation. In the remainder of this section, we report two interesting cases that passed in both validations and the reasons why branches were invalidated.

\noindent\textit{$>$ Successful interesting cases}

\vspace{5pt}
\noindent\textit{The bug-triggering test case already existed in the first passing build in case \#2.} When we defined the case \#2 (see \autoref{sec:build-pair-cases}), our hypothesis was that the first passing build contains a bug, but it passed in Travis because there was no test case to expose the bug. However, during the manual validation, we found a few cases where the bug-triggering test case already existed in the first passing build, but it was marked to be ignored when running the test suite. This is the case, for instance, of the pair of builds \href{https://travis-ci.org/raphw/byte-buddy/builds/352481508}{352481508}--\href{https://travis-ci.org/raphw/byte-buddy/builds/352894244}{352894244} from the raphw/byte-buddy project.

\vspace{5pt}
\noindent\textit{The commit message hides a bug fix.} The commit message is sometimes unclear on the changes performed in a commit. We found an interesting case that the \bearsCollector reproduced, where several test cases failed due to \texttt{NullPointerException} when reproducing the $b_{buggy}$ candidate. The message of the commit that triggered $b_{patched}$ candidate, however, was ``Refactor tests''. The source code diff between the commits that triggered both builds really suggested a fix for \texttt{NullPointerException}, so we contacted the developer in the same day that the bug fix commit happened (this pair was collected in the real time execution). The developer confirmed the existence of the bug fix, and that he was in the middle of a bug-hunting. This was the case of the pair of builds \href{https://travis-ci.org/vitorenesduarte/VCD-java-client/builds/437204853}{437204853}--\href{https://travis-ci.org/vitorenesduarte/VCD-java-client/builds/437571024}{437571024} from the vitorenesduarte/VCD-java-client project.

\vspace{5pt}
\noindent\textit{$>$ Invalid branches during the manual validation}

\vspace{5pt}
\noindent\textit{Refactoring/cleaning.} We obtained several refactorings and cleanings from pairs of builds in case \#2 (passing-passing builds). For instance, the pair of builds \href{https://travis-ci.org/aicis/fresco/builds/330415018}{330415018}--\href{https://travis-ci.org/aicis/fresco/builds/330418847}{330418847} from the aicis/fresco project is related to a cleaning instead of a bug fix. A part of the source code was removed, and the test code was adapted so the tests pass on the cleaned code, but not on the previous version of the code without the cleaning.

\vspace{5pt}
\noindent\textit{Feature addition/enhancement.} We also obtained feature addition from pairs of builds in case \#2 (passing-passing builds). The changes in the tests in the second passing build are either to test the new feature or adapted to work properly on the source code containing the new feature, thus changed tests fail when executed on the previous version of the code without the feature. For instance, the pair of builds \href{https://travis-ci.org/traccar/traccar/builds/217413927}{217413927}--\href{https://travis-ci.org/traccar/traccar/builds/217431352}{217431352} from the traccar/traccar project is related to a feature addition, where the commit associated with the second build contains the message ``Add new Aquila protocol format''.

\begin{table*}[t]
    \caption{Excerpt of open-source projects contained in the \bearsBenchmark. The diversity of domains, age and size is higher than Defects4J and Bugs.jar.}
    \label{tab:projects}
    \centering
    \begin{threeparttable}[b]
    \begin{tabular}{p{.21\textwidth} c p{.24\textwidth}|c r r|r r r}
        \toprule
        \multicolumn{3}{c|}{Project} & \multicolumn{3}{c|}{GitHub info\tnote{a}} & \multicolumn{3}{c}{\bears info\tnote{b}} \\
        Name & Type & Domain/Topics & Creation & \#Contrib. & \#Commits & LOC & \#Tests & \#Bugs \\
        \midrule
        \rowcolor{gray!15}
        INRIA/spoon & Lib & Java code analysis and transformation & 2013/11 & ~47 & 2,804 & 76,295 & 1,114 & 62 \\
        traccar/traccar & App & GPS tracking system & 2012/04 & ~87 & 5,416 & 43,397 & 255 & 42 \\
        \rowcolor{gray!15}
        {\scriptsize FasterXML/jackson-databind} & Lib & general data binding (e.g. JSON) & 2011/12 & 141 & 5,038 & 99,151 & 1,711 & 26 \\
        {\scriptsize spring-projects/spring-data-commons} & Lib & spring data, data access & 2010/11 & ~62 & 1,793 & 36,479 & 2,029 & 15 \\
        \rowcolor{gray!15}
        debezium/debezium & App & platform for change data capture & 2016/01 & ~69 & 1,439 & 53,318 & 508 & 7 \\
        raphw/byte-buddy & Lib & runtime code generation for the JVM & 2013/11 & ~39 & 4,640 & 140,087 & 8,066 & 5 \\
        \rowcolor{gray!15}
        {\scriptsize SzFMV2018-Tavasz/AutomatedCar} & App & passenger vehicle behavior simulator & 2018/01 & ~33 & 862 & 2,197 & 48 & 2 \\
        rafonsecad/cash-count & App & accounting software back-end & 2018/09 & 1 & 29 & 759 & 16 & 2 \\
        \rowcolor{gray!15}
        Activiti/Activiti & App & business process management & 2012/09 & 160 & 8,041 & 205,097 & 1,952 & 1 \\
        pippo-java/pippo & Lib & micro Java web framework & 2014/10 & ~22 & 1,341 & 19,738 & 131 & 1 \\
        \bottomrule
    \end{tabular}
    \begin{tablenotes}
        \item[a]{The number of contributors and the number of commits were collected in December 21, 2018.}
        \item[b]{The metrics LOC and number of tests are averaged over the collected bugs.}
    \end{tablenotes}
    \end{threeparttable}
\end{table*}

\vspace{5pt}
\noindent\textit{Duplicate patches.} We obtained several branches with duplicate patches. When investigating the reason why this happens, we found the following scenario. Some changes (such as bug fix) were applied in a branch $X$ in a given project, and we obtained a \bears branch from the pair of builds related to the performed changes in the branch $X$. Then, these changes were also applied to a branch $Y$ in the given project, and we obtained a \bears branch from a pair of builds in the branch $Y$ too. As a result, the two \bears branches happen to contain the same patch. For instance, we obtained branches from the pairs \href{https://travis-ci.org/FasterXML/jackson-databind/builds/176912167}{176912167}--\href{https://travis-ci.org/FasterXML/jackson-databind/builds/190405643}{190405643} and \href{https://travis-ci.org/FasterXML/jackson-databind/builds/183487103}{183487103}--\href{https://travis-ci.org/FasterXML/jackson-databind/builds/190406891}{190406891} from the FasterXML/jackson-databind project. The changes performed in the commit that triggered \href{https://travis-ci.org/FasterXML/jackson-databind/builds/190405643}{190405643} in the branch ``2.7'' were merged into the branch ``2.8'', creating a new commit and triggering \href{https://travis-ci.org/FasterXML/jackson-databind/builds/190406891}{190406891}.

\vspace{5pt}
\noindent\textit{Unrelated commits combined.} We obtained branches that store patches containing unrelated changes from multiple commits combined due to pairs of builds that are not from immediately subsequent commits.
We observed two cases where this may happen.
In the first one, the developer does more than one commit locally and push them to the remote repository at once. In this case, only the last pushed commit triggers a build in Travis, meaning that there are commits with no associated build between the build from the last commit and its previous build.
In the second case, a build $b_{n}$ finished to run in Travis after the build $b_{n+1}$, so when obtaining the previous build from $b_{n+2}$, we obtain $b_{n}$. As in the first case, there are commits between the pair of builds. This is the case of the pair of builds \href{https://travis-ci.org/INRIA/spoon/builds/234112974}{234112974}--\href{https://travis-ci.org/INRIA/spoon/builds/234114955}{234114955} from the INRIA/spoon project, where two builds were triggered between the pair of builds.

\vspace{5pt}
\noindent\textit{Bug fix including other changes.} Finally, we discarded some branches containing genuine bug fixes, but that also included other changes such as refactoring and code formatting. An example of a discarded branch is from the pair of builds \href{https://travis-ci.org/molgenis/molgenis/builds/371024842}{371024842}--\href{https://travis-ci.org/molgenis/molgenis/builds/371501238}{371501238} from the molgenis/molgenis project. We do not claim that the \bearsBenchmark contains only branches with isolated bug fixes, but we avoided the inclusion of branches containing bug fix mixed with other changes to make easier studies on patches such as in \cite{Sobreira2018defects4J-dissection}.

\vspace{5pt}
\noindent\textit{$>$ Invalid branches during the automatic validation}

\vspace{5pt}
The automatic validation, which runs checks in Travis on the content of branches (from pull requests in the \bearsBenchmark repository), invalidated branches mainly because a failure happened when validating the patched program version by building and running tests, which suggests, for instance, the existence of flaky tests.

\section{Content of the \bearsBenchmark}\label{sec:content-of-bears}

The collection of bugs presented in the previous section resulted in \nbBearsBugs real, reproducible bugs, which constitute the version 1.0 of the \bearsBenchmark. Out of \nbBearsBugs, 19 bugs are from builds in case \#1 (failing-passing builds) and 232 bugs are from builds in case \#2 (passing-passing builds with test changes). In this section, we present general information on these bugs.

\subsection{Constituent Projects}

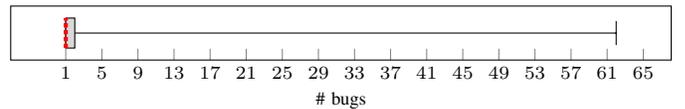
\begin{figure}[h!]
  \centering
  \scriptsize
  \begin{tikzpicture}
    \begin{axis}
      [width=1.17\linewidth,
      height=2.3cm,
      enlarge y limits=0.4,
      xlabel=\# bugs,
      xlabel near ticks,
      xtick={1,5,...,65},
      xtick pos=left,
      ytick=\empty]
      \addplot[draw=black,fill=black!15,
      boxplot prepared={
        lower whisker=1,
        lower quartile=1,
        median=1,
        upper quartile=2,
        upper whisker=62,
        every median/.style={densely dotted,red,ultra thick},
      }]
      table [row sep=\\,y index=0] { \\ };
    \end{axis}
  \end{tikzpicture}
  \vspace{-12pt}
  \caption{Distribution of the number of bugs per project.}
  \label{fig:distribution-bugs-per-project}
\end{figure}

The \nbBearsBugs bugs come from \nbBearsProjects projects. \autoref{fig:distribution-bugs-per-project} shows the distribution of the number of bugs per project. Note that a few projects have a high number of bugs (the maximum is a project with 62 bugs), and there is a high concentration of projects with one (the median) or a few bugs. This high concentration is mainly due to the real time collection (execution round \#3).

\autoref{tab:projects} presents ten out of the \nbBearsProjects projects. The first five projects are the ones with more bugs in the \bearsBenchmark, and the other five were randomly selected. For each project, we present its type (library or application), domain, creation date on GitHub, the number of contributors and commits, and two size metrics (LOC and \# Tests) averaged over the collected bugs.

From this table, we note that the bugs in the \bearsBenchmark are from at least ten different project domains. Moreover, the projects considerably differ in age and size: there are bugs from a 8-years old project (spring-data-commons) and also from recent projects less than 1-year old (AutomatedCar and cash-count). Considering size, there are bugs from projects ranging from 759 LOC to 205 KLOC, with number of tests ranging from 16 to 8K. Overall, this shows that the bugs in the \bearsBenchmark are from projects that are diverse in many different aspects.

\subsection{Constituent Exception Types}

The \bearsBenchmark is a benchmark of reproducible bugs.
Per our process, the \bearsCollector runs JUnit test suites and collects test failures. JUnit reports non-passing test cases in two different ways: test failure and test in error.
JUnit reports a test failure when an asserted condition in a test case does not hold~\cite{Langr2015}, i.e. when an expected result value does not match the actual value. On the other hand, JUnit reports a test error when an exception is thrown and not caught during the execution of the test~\cite{Langr2015}, e.g. an array index out of bounds.

During the reproduction process, the \bearsCollector extracts the information on test failure and test in error from the reproduced bug. This type of information is useful for automatic program repair research: for instance, NPEFix~\cite{Durieux2017-NPEFix} focuses on repairing bugs that are exposed by null pointer exception.
In the \bearsBenchmark, the top-2 most occurring exceptions are \texttt{AssertionError} (108 bugs) and \texttt{ComparisonFailure} (31 bugs), both test failures, followed by \texttt{NullPointerException} (26 bugs), one of the most frequent runtime exceptions.

\subsection{Constituent Patches}

The \bearsBenchmark stores the buggy and patched program versions for each bug. The diff between these two versions contains the patch created by a developer to fix the bug. To provide a preliminary view on the patches included in the \bearsBenchmark, we calculated their size in number of lines and their spreading in number of files.

To calculate patch size, we sum the number of added, deleted, and modified lines (sequences of added and deleted lines). To calculate patch spreading, we count the number of modified files.
\autoref{tab:patch-size-and-spreading} shows the results on both metrics. Note that the patches involve one to 312 lines, and 50\% of the patches involve at most eight lines. The majority of patches (75\%) change at most a single file, and considering all patches, at maximum ten files are modified.

\begin{table}[h]
    \caption{Descriptive statistics on the size and spreading of the patches included in the \bearsBenchmark.}
    \label{tab:patch-size-and-spreading}
    \centering
    \begin{tabular}{l r r r r r}
        \toprule
        {} & Min & 25\% & 50\% & 75\% & Max \\
        \midrule
        Patch size (lines) & 1 & 3 & 8 & 18.5 & 312 \\
        Patch spreading (files) & 1 & 1 & 1 & 1 & 10 \\
        \bottomrule
    \end{tabular}
\end{table}

\section{Discussion}\label{sec:discussion}

To date, we are the first to report on a bug collection process based on Continuous Integration. The issues we had during the reproduction attempts (e.g. 33.6\% of them failed when building) are valuable insights to those who are interested in researching on  bug collection. In this section, we present challenges and limitations when designing \bears, developing the \bearsCollector, and creating the \bearsBenchmark.

\subsection{Challenges}

The development of the \bearsCollector was a challenge itself. This is mainly due to the high automation required to collect and store bugs. The different steps must be integrated, so that given a configuration for the process, the \bearsCollector scans builds, performs reproduction attempts, and stores the successful ones in a publicly available GitHub repository, with a standard organization and a proper metadata file (\texttt{bears.json}) containing information on the reproduction.

A more specific challenge we faced is due to case \#2 (passing-passing builds with test changes) from multi-module projects and non-standard single-module projects. In case \#2, the source code and test code files need to be identified, so that they can be mixed for testing the source code of the first passing build with the tests from the second passing build. In a standard single-module project, the source code files are maintained in the folder \texttt{src/main/java}, and the test files are maintained in the folder \texttt{src/test/java}, both folders localized in the root folder of the project. However, in multi-module projects, each module has its own organization, and in non-standard single-module projects, the paths of the folders are different from the standard ones. Thus, for case \#2, the checkout of source code and test code folders was a concern. To overcome this, given a multi-module project or non-standard single-module project, we search in all folders of the project for \texttt{pom.xml} files, and we try to compute the source code and test code folders throughout the given paths.

\subsection{Limitations}


The \bearsCollector only covers Maven projects: the core steps of the reproduction process were implemented specifically for Maven projects. These steps, in fact, rely on Maven to build the project and to run tests. However, the approach itself is independent from the used build tool. Additional work is needed to be able to collect bugs from Travis CI builds that are not from Maven projects (e.g. from Gradle projects).

The environment used in the reproduction process by the \bearsCollector to build and run tests on program versions is a different environment from Travis CI.
The environment used by Travis can be configured by the developers, ranging from OS to customized scripts.
The environment used by the \bearsCollector is the same for every project.
For that reason, the \bearsCollector might fail in reproduction attempts, as well as producing false positive bugs.

An open challenge is the manual validation. Like the state-of-the-art Java benchmarks, a manual analysis is always done before adding a given bug in the benchmark. The manual analysis is a difficult and time-consuming task. Some candidates are simple to validate, for instance, when a supposed bug fixing commit contains a reference to an issue in the repository of the project, describing the issue addressed in the commit. However, other ones are harder to validate, since the analysis of the source code diff of the two program versions is required. This manual step is a bottleneck to scale the process up.

\subsection{Threats to Validity}

As any tool, the \bearsCollector is not free of bugs. A bug in the \bearsCollector might impact the results reported on the execution rounds we performed (presented in \autoref{sec:collecting-bugs}). However, the \bearsCollector is open-source and publicly available for other researchers and potential users.

Bug candidates might be wrongly validated before their inclusion in the \bearsBenchmark. For this first version of the \bearsBenchmark, the manual analysis (presented in \autoref{sec:validating-branches}) was performed by the first author of this paper. Despite of her careful manual analysis, any manual work is prone to mistakes. Our intention is, however, to create an open environment for contributions, where one might 1) flag possibly incorrect branches added in the \bearsBenchmark, and 2) participate in the manual validation when pull requests are opened in the \bearsBenchmark. Additionally, bugs triggered by flaky tests might exist in the \bearsBenchmark. However, each bug was reproduced twice in the same environment (i.e. in the reproduction and automatic validation phases), which mitigates the threat of having bugs with flaky tests.

\section{Related Work}\label{sec:related-work}

Benchmarks of bugs are assets that have been used in software bug-related research fields to support empirical evaluations.
Several benchmarks were first created for the software testing research community, such as Siemens~\cite{Hutchins1994} and SIR~\cite{Do2005}, two notable and well-cited benchmarks. The majority of bugs in these two benchmarks were seeded in existing program versions without bugs, which is farther away from \bears that targets real bugs.

BugBench~\cite{Lu2005} contains real bugs, and aims to support evaluations on bug detection tools. A major difference between BugBench and the \bearsBenchmark is that BugBench was mainly built by manual effort, including the creation of tests to trigger bugs.
iBugs~\cite{Dallmeier2007}, on the other hand, was created by automatically identifying bug fixes from the history of a project, by searching log messages for references to bug reports. iBugs does not include bug-triggering test for all bugs, which meets one of its purpose (static bug localization tools), but it is not suitable for automatic program repair.

To the best of our knowledge, the first benchmarks proposed for automatic program repair research are ManyBugs and IntroClass~\cite{LeGoues2015}. ManyBugs contains 185 bugs collected from nine large, popular, open-source programs. On the other hand, IntroClass targets small programs written by novices, and contains 998 bugs collected from student-written versions of six small programming assignments in an undergraduate programming course. Both benchmarks are for the C language.

More recently other benchmarks were proposed for automatic program repair. Codeflaws~\cite{Tan2017} contains 3,902 bugs extracted from programming contests available on Codeforces. Codeflaws is also for the C language, and the programs range from one to 322 lines of code. QuixBugs~\cite{Lin2017} is a multi-lingual benchmark, which contains single line bugs from 40 programs translated to both Java and Python languages.

The closest benchmarks to \bears are Defects4J~\cite{Just2014} and Bugs.jar~\cite{Saha2018}, both for Java. Defects4J contains 395 reproducible bugs collected from six projects, and Bugs.jar contains 1,158 reproducible bugs collected from eight Apache projects. To collect bugs, the approach used for both benchmarks is based on bug tracking systems, and they contain bugs from large, mature projects. \bears, on the other hand, was designed to collect bugs from a diversity of projects other than large and mature ones: we break the need of projects using bug tracking systems. Note that bug tracking systems are used in the direction of documenting bugs. Continuous Integration, on the other hand, is used to actually build and test a project, which is closer to the task of identifying reproducible bugs.

\section{Conclusion}\label{sec:conclusion}

In this paper, we have presented the \bears project, including its approach to collect and store bugs (the \bearsCollector) and a collection of \nbBearsBugs bugs from \nbBearsProjects projects (the first version of the \bearsBenchmark). The \bearsCollector performs attempts to reproduce buggy and patched program versions from builds in Travis CI: builds can be scanned from specific projects and a given time window, or from any project in real time.

We designed \bears to maximize the automation for collecting bugs, in order to maximize diversity of projects.
We have also made it extensible, for the research community to contribute with new bugs to be added in the \bearsBenchmark. Our intention, if not dream, is a community-driven benchmark, where time and effort is spent by the community to minimize all kinds of bias.

One can find the artifacts produced by this work in the links presented below:

\begin{center}
    \footnotesize
    \begin{tabular}{p{.14\textwidth} l}
        \toprule
        \bearsBenchmark & \url{https://github.com/bears-bugs/bears-benchmark} \\
        \bearsCollector & \url{https://github.com/bears-bugs/bears-collector} \\
        Bug/patch browser & \url{https://bears-bugs.github.io/bears-benchmark} \\
        Data from \autoref{sec:collecting-bugs} & \url{https://github.com/bears-bugs/saner2019-data} \\
        \bottomrule
    \end{tabular}
\end{center}

Future work can be carried out in two different directions: improving the \bearsCollector and enhancing the \bearsBenchmark. For the former, one future work is to create heuristics to discard \bears branches that do not look like to contain a bug fix (e.g. branches containing refactoring): this would greatly minimize the effort spent in the manual validation. For the latter, studies on the characteristics of the bugs and their patches, such as repair patterns, should be conducted. Finally, the collection of bugs in real time opens the opportunity of contacting the developers who have just fixed bugs, which would be an invaluable source of information about the bugs as well as the bug fixing activity.

\section*{Acknowledgment}

We acknowledge the Brazilian National Council for Scientific and Technological Development--CNPq, CAPES, and FAPEMIG for partially funding this research. Fernanda Madeiral would also like to thank Thomas Durieux for discussions around \bears, and the Spirals team/INRIA for the resources used for the collection of bugs.

\clearpage

\bibliographystyle{IEEEtran}
\bibliography{references}

\end{document}